\documentclass{article}
\usepackage{sw20spie}
\usepackage{graphicx}
\usepackage{amsmath,amssymb,bbm}
\usepackage{color}

\begin{document}

\title{{\bf $\pi_{e3}$ form factor $f_{-}$ near the mass shell}}
\author{ { \Large M. I. Krivoruchenko} \vspace{15pt} \\
%EndAName
Institute for Theoretical and Experimental Physics$\mathrm{,}$
B. Cheremushkinskaya 25\\
117218 Moscow, Russia\\
Moscow Institute of Physics and Technology$\mathrm{,}$
141700 Dolgoprudny$\mathrm{,}$ Russia, and \\
Bogoliubov Laboratory of Theoretical Physics$\mathrm{,}$ Joint Institute for Nuclear Research\\
141980 Dubna$\mathrm{,}$ Russia
}
\vspace{50pt}
\date{}

\maketitle

%\begin{center}
%{\bf \large Abstract}
%\end{center}

\begin{abstract}
The generalized Ward-Takahashi identity (gWTI) in the pion sector for broken isotopic symmetry is derived
and used for the model-independent calculation of the longitudinal form factor $f_{-}$ of the $\pi_{e3}$ vector vertex.
The on-shell $f_{-}$ is found to be proportional to the mass difference of the pions and the difference between the
vector isospin $T = 1$ and scalar isospin $ T = 2 $ pion radii.
A numerical estimate of the form factor yields a value two times higher than the previous estimate from the quark model.
%%%%%%%%%%%%%%%%%%%%%%%%%%%%%%%%%%%%%%%%%%%%%
Off-shell form factors are known to be ambiguous because of the gauge dependence and the freedom in the parameterization of the fields.
The near-mass-shell $f_{-}$ appears to be an exception, allowing for experimental verification of the consequences of the gWTI .
We calculate the near-mass-shell $f_{-}$ using the gWTI and dispersion techniques.
The results are discussed in the context of
%the condition of the conservation of vector current.
the conservation of vector current (CVC) hypothesis.
\end{abstract}

%\vskip 90pt
%\vskip 40pt
%\baselineskip=24pt
\pagebreak
%\tableofcontents
\setcounter{page}{1}

%%%%%%%%%%%%%%%%%%%%%%%%%%%%%%%%%%%%%%%%%%%%%%%%%%%%%%%%%%%%%%%%%%%%%%%%
%%%%%%%%%%%%%%%%%%%%%%%%%%%%%%%%%%%%%%%%%%%%%%%%%%%%%%%%%%%%%%%%%%%%%%%%
\section{Introduction}
%%%%%%%%%%%%%%%%%%%%%%%%%%%%%%%%%%%%%%%%%%%%%%%%%%%%%%%%%%%%%%%%%%%%%%%%
%%%%%%%%%%%%%%%%%%%%%%%%%%%%%%%%%%%%%%%%%%%%%%%%%%%%%%%%%%%%%%%%%%%%%%%%

The $\pi^{+} \to \pi^{0} e^{+} \nu_{e}$ decay ($\pi_{e3}$) is one of the main semileptonic electroweak processes. The vector nature of the transition, its simple kinematics, and the precise measurement of the partial width make this decay particularly attractive for testing the Standard Model.

The decay amplitude is proportional to the $V_{ud}$ element of the Cabibbo-Kobayashi-Maskawa (CKM) matrix.
Radiative corrections and pion structure effects in the $\pi_{e3}$ decay have been calculated with high accuracy \cite{Sirl1978,Sirl1994,Pich2003,Pich2003a,Jaus1999,Jaus2001}, sufficient for verification of the unitarity of the CKM matrix.
The experimental data, however, are not yet sufficiently precise for this purpose \cite{PIBETA,Poca2010}.

Measurements of the $\pi_{e3}$ decay are also motivated by the possibility of testing the conservation of vector current (CVC) in the meson sector. The CVC hypothesis \cite{Gers1955,Gers1955b,Feyn1958} suggests that the isovector component of the electromagnetic current
and the charged components of the weak vector current belong to the same isospin triplet. In the limit of exact isotopic symmetry,
conservation of the electromagnetic current implies conservation of the weak vector current.

Off the mass shell, the CVC is equivalent to the Ward-Takahashi identity (WTI) for the isospin $SU(2)$ group. The WTI, however, is of greater generality and leads to useful relationships between off-shell form factors, including those that vanish when some of the external legs are on shell.
The violation of isotopic symmetry, %that is
associated with the small mass difference between the up and down quarks and the electromagnetic and weak interactions,
results in the non-conservation of the charge-changing components of the weak vector current.
For the broken symmetry, the CVC condition and the WTI are replaced by partial CVC and the generalized WTI (gWTI),
which are especially
%particularly
sensitive to the pattern of isotopic symmetry breaking.

Off-shell form factors enter the descriptions of nucleon knockout reactions \cite{Tiem1990,BosJ1993} and bremsstrahlung reactions \cite{Habe2012}. As a result of shifting away from the mass shell, the electromagnetic currents of bound nucleons are different from the free currents, which also leads to observable effects \cite{Schw2001}.

The parameterization of the degrees of freedom associated with the pion field can be performed in various ways, which produces off-shell ambiguity in the amplitudes. The on-shell form factors are related to the asymptotic states and are uniquely defined. This statement is known as the equivalence theorem (ET) \cite{Haag1958,Chis1961,Kame1961}.
The off-shell form factors, although they contribute to the physical amplitudes, depend on this parameterization and cannot be measured experimentally \cite{Fear2000,Fear2000a}.

We report a notable exception to this rule.
The longitudinal part of the $\pi_{e3}$ vertex is shown to be uniquely defined near the mass shell and
fundamentally accessible to measurements, and therefore, it can be investigated for its consistency with the gWTI.

In quark models, pions are usually poorly described because of their Goldstone boson nature.
As an application of the gWTI, we derive a model-independent expression and provide a numerical estimate for the longitudinal form factor $f_{-}$.
The structure of this paper is as follows. Sect. 2 begins with a discussion of the constraints imposed by the WTI of the symmetry group $U(1)$ on the electromagnetic pion form factors. In Sect. 3, a generalization of the WTI for broken isotopic symmetry is derived, and in Sect. 4, it is used to find the relationships between weak vector form factors. The near-mass-shell form factor $f_{-}$ is then expressed in terms of the pion mass difference and the pion radii in the isospin $T = 1$ and $2$ channels. In Sect. 5, we focus on extracting the isospin $T=2$ pion radius from the experimental data on $\pi\pi$ scattering phase shifts and numerically evaluate $f_{-}$. Finally, we conclude in Sect. 6 with a summary and a discussion.

%%%%%%%%%%%%%%%%%%%%%%%%%%%%%%%%%%%%%%%%%%%%%%%%%%%%%%%%%%%%%%%%%%%%%%%%%%%%%%%%%%%%%%%%%%%%%%%%%%%%%%%%%%%%%%%%%%%%%%%%%%%%%%%%%%%%%%%%%%%%%%%%%%%%%%
\section{U(1) vector vertex}
\renewcommand{\theequation}{2.\arabic{equation}}
\setcounter{equation}{0}
%%%%%%%%%%%%%%%%%%%%%%%%%%%%%%%%%%%%%%%%%%%%%%%%%%%%%%%%%%%%%%%%%%%%%%%%%%%%%%%%%%%%%%%%%%%%%%%%%%%%%%%%%%%%%%%%%%%%%%%%%%%%%%%%%%%%%%%%%%%%%%%%%%%%%%

The on-shell conserved vector current of a charged scalar particle is determined by one form factor. Off the mass shell, there are two form factors. In the most general case, the current can be written as follows
\begin{equation}
\Gamma_{\mu }(p^{\prime},p) = (p^{\prime} + p)_{\mu}{\mathcal{F}}%
_{1} + q_{\mu}(p^{\prime 2} - p^{2}){\mathcal{F}}_{2},
\label{1}
\end{equation}
where $q=p^{\prime }-p$ is the momentum transfer. The form factors
${\mathcal{F}}_{i}$ are symmetric functions of $p^{\prime 2}$ and $p^{2}$ and arbitrary functions of $q^{2}$ and the physical mass $m$ of the charged pion.
The factor $p^{\prime 2}-p^{2}$ in the second term is added to ensure the negative $C$ parity of the current.
The form factor decomposition (\ref{1}) arises in scalar quantum electrodynamics (QED) (see, e.g., \cite{Ball1980}) and in chiral perturbation theory ($\chi$PT) (see, e.g., \cite{Rudy94}). The WTI of the $U(1)$ symmetry group establishes a relationship between ${\mathcal{F}}_{1}$ and ${\mathcal{F}}_{2}$ that is identical to that of Eq.~(\ref{6a}).

Isotopic symmetry implies that the mass difference of non-strange quarks can be neglected and that the electromagnetic and weak interactions are switched off beyond tree level. In this limit, the weak interaction is described by an isovector current $\Gamma _{\mu }^{a}(p^{\prime },p)=T^{\alpha }\Gamma _{\mu}(p^{\prime },p)$, where the $T^{\alpha }$ are isospin generators. As a result, exact CVC occurs: a dressed vertex forms an isospin triplet, and the weak vector current is conserved.

The WTI associated with the $U(1)$ symmetry group imposes a constraint on the vertex (see, e.g., \cite{Bjor1965}):
\begin{equation}
\Delta ^{-1}(p^{\prime })-\Delta^{-1}(p) = q_{\mu }\Gamma _{\mu }(p^{\prime },p),  \label{2}
\end{equation}
where $\Delta (p)=p^{2}-m^{2}-\Sigma (p^{2},m)$ is the renormalized pion
propagator. The self-energy operator satisfies%
\begin{eqnarray}
\Sigma (m^{2},m) &=&0,  \label{5} \\
\left. \frac{\partial }{\partial p^{2}}\Sigma (p^{2},m)\right\vert
_{p^{2}=m^{2}} &=&0.  \label{6}
\end{eqnarray}%
Combining Eq.~(\ref{2}) with Eq.~(\ref{1}) yields
\begin{equation}
q^{2}\mathcal{F}_{2}(p^{\prime 2},p^{2},q^{2})=\frac{\Delta ^{-1}(p^{\prime
})-\Delta ^{-1}(p)}{p^{\prime 2}-p^{2}} - \mathcal{F}_{1}(p^{\prime
2},p^{2},q^{2}).  \label{6a}
\end{equation}
In the limit $p^{\prime }{}^{2}=p^{2}=m^{2}$, we obtain
\begin{equation}
{\mathcal{F}}_{2}(m^{2},m^{2},q^{2})=\frac{1-{\mathcal{F}}%
_{1}(m^{2},m^{2},q^{2})}{q^{2}}.  \label{7}
\end{equation}%
In the vicinity of $q^{2}=0$, the form factor ${\mathcal{F}}_{1}$ can be expanded to give
\begin{equation}
{\mathcal{F}}_{2}(m^{2},m^{2},0)=-\frac{1}{6}\left\langle
r^{2}\right\rangle _{v},  \label{8}
\end{equation}%
where $\left\langle r^{2}\right\rangle _{v}$ is the vector charge radius.

The pion form factor ${\mathcal{F}}_{1}(p^{\prime 2},p^{2},q^2)$, which describes strong interaction effects, is an analytic function of the complex variable $p^2$ at least in the circle $|p^2 - m^2| < 8m^2$. The nearest to the mass shell singularity at $p^2 = 9m^2$ is caused by the three-pion threshold. A Taylor series in powers of $p^2 - m^2$ converges within that circle. The same is true for the variable $p^{\prime 2}$. A Taylor series in powers of $q^2$ converges for $|q^2| < 4m^2$, where the radius of convergence is determined by the two-pion threshold.

The equivalence of the Coulomb and Lorentz gauges in QED was rigorously proved in Refs. \cite{Bial1970,Hoof1972}. On shell, the amplitudes are gauge-invariant, whereas the off-shell dependence on the gauge persists.
${\mathcal{F}}_{1}$ and, by virtue of Eq.~(\ref{7}), ${\mathcal{F}}_{2}$ are thus uniquely defined, when both legs of the charged pion are on the mass shell. To first order in the displacement from the mass shell, the longitudinal component of the vertex is also gauge invariant, as is evident from Eq.~(\ref{1}). Aside from these cases, $\Gamma _{\mu }(p^{\prime },p)$ depends on the gauge and cannot be measured.

${\mathcal{F}}_{2}$ with two on-shell legs
$p^{\prime 2} = p^{2} = m^2$ does not contribute to the current. Off the mass shell, however, ${\mathcal{F}}_{2}$ does contribute, and its contribution is uniquely determined by the WTI.
Isotopic rotation
of ${\mathcal{F}}_{2}$ is not sufficient to obtain a full weak-interaction vertex. We show that the violation of isotopic symmetry generates an isospin $T = 2$ contribution that is unrelated to isotopic rotation.

%%%%%%%%%%%%%%%%%%%%%%%%%%%%%%%%%%%%%%%%%%%%%%%%%%%%%%%%%%%%%%%%%%%%%%%%%%%%%%%%%%%%%%%%%%%%%%%%%%%%%%%%%%%%%%%%%%%%%%%%%%%%%%%%%%%%%%%%%%%%%%%%%%%%%%
\section{Generalized Ward-Takahashi identity
for broken isotopic symmetry}
\renewcommand{\theequation}{3.\arabic{equation}}
\setcounter{equation}{0}
%%%%%%%%%%%%%%%%%%%%%%%%%%%%%%%%%%%%%%%%%%%%%%%%%%%%%%%%%%%%%%%%%%%%%%%%%%%%%%%%%%%%%%%%%%%%%%%%%%%%%%%%%%%%%%%%%%%%%%%%%%%%%%%%%%%%%%%%%%%%%%%%%%%%%%

In order to find the longitudinal form factor of the weak vector current, we derive a generalization of the WTI that is associated with the replacement of the exact $U (1)$ symmetry by the broken $SU(2)$ symmetry.

Let us consider the variation of the pion propagator
\[
i\Delta ^{\alpha \beta }(x^{\prime },x)=\langle 0|T\varphi ^{\alpha
}(x^{\prime })\varphi ^{\beta }(x)|0\rangle
\]
under the $SU(2)$ transformation
\begin{equation}
\varphi \rightarrow \varphi ^{\prime }=e^{-i\chi }\varphi .
\label{21}
\end{equation}
In terms of a Cartesian basis, the pion field $\varphi^{\alpha}(x)$ is Hermitian, and $\chi$ is an infinitesimal imaginary antisymmetric matrix that can be expanded in the group generators with real coefficients $\chi = \sum_{a} \chi^{a} T^{a}$. In the conventional normalization, $\mathrm{Tr}\left( T^{a}T^{b}\right) =2\delta ^{ab}$. The variation of the pion propagator is given by
\begin{equation}
i\delta \Delta (x^{\prime },x)=\chi (x^{\prime })\Delta (x^{\prime
},x)-\Delta (x^{\prime },x)\chi (x).
\label{21b}
\end{equation}
In matrix notation, $\Delta ^{-1}i\delta \Delta \Delta ^{-1}=-i\delta \Delta^{-1}=[\Delta ^{-1},\chi ]$.

Independent calculation of $\delta \Delta $ involves the effective Lagrangian, which we consider to be of the form
\begin{equation}
\mathcal{L}_{\mathrm{eff}}=\mathcal{L}_{\mathrm{eff}}((D_{\mu }\varphi
)^{\dagger },D_{\mu }\varphi ,\varphi ^{\dagger },\varphi ),  \label{eL}
\end{equation}%
%%%%%%%%%%%%%%%%%%%%%%%%%%%%%%%%%%%%%%%%%%%%%%%%%%%%%%%%%%%%%%%%%%%%%%%%%%%%%%%%%%%%%%%%%%%%%%%%%%%%%%%%%%%%%%%%%%%%%%%%%%%%%%%%%%%%%%%%%%%%%%%%%%%%%%
\begin{figure} [t]
\begin{center}
\includegraphics[width=0.85\textwidth]{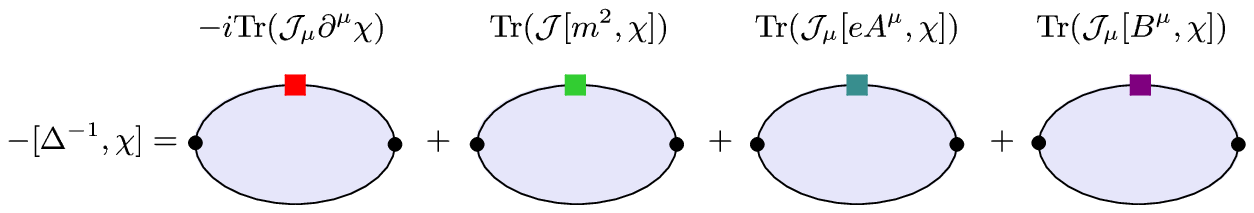}
\caption{Schematic representation of the gWTI for the pion propagator with broken isotopic symmetry. The blobs denote the dressed vertices. The first one is a Lorentz vector entering the pion $\beta$ decay. The last three are Lorentz scalars; they are responsible for the symmetry breaking by the pion masses and the electromagnetic and weak interactions.}
\label{fig1}
\end{center}
\end{figure}
%%%%%%%%%%%%%%%%%%%%%%%%%%%%%%%%%%%%%%%%%%%%%%%%%%%%%%%%%%%%%%%%%%%%%%%%%%%%%%%%%%%%%%%%%%%%%%%%%%%%%%%%%%%%%%%%%%%%%%%%%%%%%%%%%%%%%%%%%%%%%%%%%%%%%%
\noindent
where $iD_{\mu }=i\partial _{\mu }-eA_{\mu }-B_{\mu }$, $A_{\mu }$ is the
electromagnetic field, $e=T^{3}$, and $B_{\mu }=B_{\mu }^{a}T^{a}$ is the weak
vector field associated with $Z^{0}$ and $W^{\pm }$ bosons.
In a Cartesian basis, $D_{\mu }$ is real. %and $\varphi ^{\dagger }=\tilde{\varphi}$.
The isotopic invariance
in $\mathcal{L}_{\mathrm{eff}}$ is broken by the mass term $\varphi^{\dagger }m^{2}\varphi$ with $[m^{2},T^{a}]\neq 0$ and by the electromagnetic and weak interactions. In other respects,
$\mathcal{L}_{\mathrm{eff}}$ is an arbitrary function, which reflects the ambiguity in the parameterization of the pion field.

The combination of $eA_{\mu} + B_{\mu}$ ensures the
implementation of the CVC condition at a bare interaction vertex: the
electromagnetic current coincides with the third component of the weak vector current by construction.
If the isotopic symmetry in $\mathcal{L}_{\mathrm{eff}}$
would not have been broken, the CVC condition could also hold for a dressed vertex, and the weak vector current could also be conserved.
In the case of this broken symmetry, however, the dressed vertex acquires an admixture of tensor components, while the charge-changing
weak vector current is no longer conserved.
The WTI can be viewed as a generalization of the CVC for an off-shell vertex. It permits generalizations for broken symmetries; thus, when the isotopic symmetry is broken, the CVC becomes partial CVC, whereas the WTI of the exact isotopic symmetry turns into the WTI of the broken isotopic symmetry.

Transformation (\ref{21}) is equivalent to the compensating transformation in $\mathcal{L}_{\mathrm{eff}}$:
\begin{equation}
\varphi \rightarrow \varphi ^{\prime }=e^{i\chi }\varphi,
\label{comp}
\end{equation}
which generates variations of the pion mass term and the vector fields in $\mathcal{L}_{\mathrm{eff}}$:
\begin{eqnarray}
m^2 &\rightarrow & m^{\prime 2} = m^2 + [m^{2},i\chi ],  \label{22a} \\
eA_{\mu } &\rightarrow &eA_{\mu }^{\prime }=eA_{\mu }+[e,i\chi ]A_{\mu },
\label{22b} \\
B_{\mu } &\rightarrow &B_{\mu }^{\prime }=B_{\mu }+\partial _{\mu }\chi
+[B_{\mu },i\chi ].  \label{22c}
\end{eqnarray}%
As a result, $\mathcal{L}_{\mathrm{eff}}$ acquires the correction
\begin{equation}
\delta \mathcal{L}_{\mathrm{eff}}=-\mathrm{Tr}\left( {\mathcal{J}}%
_{\mu }(\partial ^{\mu }\chi +[e,i\chi ]A^{\mu } \right. %\nonumber \\
\left. +[B^{\mu },i\chi ])\right) -\mathrm{Tr}\left( {\mathcal{J}}[m^{2},i\chi ]\right) ,
\label{leff}
\end{equation}
where
\begin{equation}
\mathcal{J}_{\mu _{\beta }}^{\alpha } =-\frac{\partial \mathcal{L}_{%
\mathrm{eff}}}{\partial B_{\alpha }^{\mu \beta }}, \;\;\;\;
\mathcal{J}_{\beta }^{\alpha } = -\frac{\partial \mathcal{L}_{\mathrm{eff}}}{\partial (m^{2})_{\alpha }^{\beta }}
\end{equation}
are the vector and scalar currents of the pion.

The response of the pion propagator to the compensating transformation is
%\begin{widetext}
\begin{eqnarray}
i\delta \Delta (x^{\prime },x) &=&\langle 0|T
\varphi (x^{\prime })
{\tilde \varphi}(x)i\int d^{4}y\delta \mathcal{L}_{\mathrm{eff}}(y)|0\rangle  \nonumber
\\
&=&\int d^{4}yd^{4}z^{\prime }d^{4}zi\Delta (x^{\prime },z^{\prime })\left[
-i\Gamma _{\mu }^{a}(z^{\prime },z,y)\partial ^{\mu }\chi ^{a}(y)+\Theta
^{a}(z^{\prime },z,y)\chi ^{a}(y)\right. \label{21c} \\
&&\;\;\;\;\;\;\;\;\;\;\;\;\;\;\;\;\;\;\;\;\;\;\;\;\;\;\;\;\;\;\;\;\;\;\;\;\;\; \left. +\Omega ^{a}(z^{\prime },z,y)\chi ^{a}(y)
\right] i\Delta (z,x),  \nonumber
\end{eqnarray}
with vertex functions defined by
\begin{eqnarray}
\Gamma _{\mu }^{a}(z^{\prime },z,y) &=&-\int d^{4}x^{\prime }d^{4}x\Delta
^{-1}(z^{\prime },x^{\prime })\langle 0|T\varphi (x^{\prime })\tilde{\varphi}%
(x)\mathrm{Tr}\left( \mathcal{{J}}_{\mu }(y)T^{a}\right) |0\rangle
\Delta ^{-1}(x,z), \\
\Theta ^{a}(z^{\prime },z,y) &=&-\int d^{4}x^{\prime }d^{4}x\Delta
^{-1}(z^{\prime },x^{\prime })\langle 0|T\varphi (x^{\prime })\tilde{\varphi}%
(x)\mathrm{Tr}\left( \mathcal{{J}}(y)[m^{2},T^{a}]\right) |0\rangle
\Delta ^{-1}(x,z), \\
\Omega ^{a}(z^{\prime },z,y) &=&-\int d^{4}x^{\prime }d^{4}x\Delta
^{-1}(z^{\prime },x^{\prime })\langle 0|T\varphi (x^{\prime })\tilde{\varphi}%
(x) \nonumber \\
&&\;\;\;\;\;\;\;\;\;\;\;\;\;\;\;\; \times \mathrm{Tr}\left( \mathcal{{J}}_{\mu }(y)[eA^{\mu }(y)+B^{\mu
}(y),T^{a}]\right) |0\rangle \Delta ^{-1}(x,z).
\end{eqnarray}%
Now, let us consider momentum space:
\begin{eqnarray}
(2\pi )^{4}\delta ^{4}(p^{\prime }-p-q)\Gamma _{\mu }^{a}(p^{\prime },p)
&=&\int d^{4}zd^{4}z^{\prime }d^{4}ye^{ip^{\prime }z^{\prime
}-ipz-iqy}\Gamma _{\mu }^{a}(z^{\prime },z,y), \\
(2\pi )^{4}\delta ^{4}(p^{\prime }-p-q)\Theta ^{a}(p^{\prime },p) &=&\int
d^{4}zd^{4}z^{\prime }d^{4}ye^{ip^{\prime }z^{\prime }-ipz-iqy}\Theta
^{a}(z^{\prime },z,y), \\
(2\pi )^{4}\delta ^{4}(p^{\prime }-p-q)\Omega ^{a}(p^{\prime },p) &=&\int
d^{4}zd^{4}z^{\prime }d^{4}ye^{ip^{\prime }z^{\prime }-ipz-iqy}\Omega
^{a}(z^{\prime },z,y).
\end{eqnarray}%
Combining Eqs.~(\ref{21b}) and (\ref{21c}) leads to
\begin{equation}
\Delta ^{-1}(p^{\prime })T^{a}-T^{a}\Delta ^{-1}(p) = q_{\mu }\Gamma _{\mu
}^{a}(p^{\prime },p) -\Theta ^{a}(p^{\prime },p)-\Omega ^{a}(p^{\prime },p).
\label{20}
\end{equation}%
%\end{widetext}
This equation generalizes the WTI for the broken $SU(2)$ symmetry. It is shown graphically in Fig. 1.

%%%%%%%%%%%%%%%%%%%%%%%%%%%%%%%%%%%%%%%%%%%%%%%%%%%%%%%%%%%%%%%%%%%%%%%%%%%%%%%%%%%%%%%%%%%%%%%%%%%%%%%%%%%%%%%%%%%%%%%%%%%%%%%%%%%%%%%%%%%%%%%%%%%%%%
\section{SU(2) vector vertex}
\renewcommand{\theequation}{4.\arabic{equation}}
\setcounter{equation}{0}
%%%%%%%%%%%%%%%%%%%%%%%%%%%%%%%%%%%%%%%%%%%%%%%%%%%%%%%%%%%%%%%%%%%%%%%%%%%%%%%%%%%%%%%%%%%%%%%%%%%%%%%%%%%%%%%%%%%%%%%%%%%%%%%%%%%%%%%%%%%%%%%%%%%%%%

In a Cartesian basis, $\Delta (x,y)=\tilde{\Delta}(y,z)$, which means that $\Delta (p)=\tilde{\Delta}(-p)$. Given that $\Delta ^{-1}(p)=
p^{2}-m^{2}-\Sigma (p^{2})$, the pion propagator is symmetric under transposition of the isospin indices: $\Delta (p)=\tilde{\Delta}(p)$.

The isospin content of a $3 \times 3$ matrix $\mathcal{M}$ can be determined using projection operators:
\begin{eqnarray}
({\mathcal{M}})^{T=0}&=&\frac{1}{2}\mathrm{Tr}[{\mathcal{M}}],  \nonumber \\
({\mathcal{M}})^{T=1}_{a}&=&\frac{1}{2}\mathrm{Tr}[{\mathcal{M}}T^{a}], \label{proj} \\
({\mathcal{M}})^{T=2}_{ab}&=&\frac{1}{2}\mathrm{Tr}
[{\mathcal{M}}(T^{a}T^{b}+T^{b}T^{a}-\frac{2}{3}\delta ^{ab}T(T+1))]. \nonumber
\end{eqnarray}
Among symmetric matrices, there are only isospin-0 or isospin-2 matrices that are proportional to the unit matrix or to quadratic forms of $T^{a}$.
The allowed pion mass operator is therefore $m^{2}+\Sigma (m^{2}) = C_{0}+C_{1}(T^{3})^{2}$, where the $C_{i}$ are numbers. The charged pions are degenerate in mass; therefore, $C$ invariance holds.

The vertex functions obey
\begin{eqnarray}
\Gamma _{\mu }^{a}(p^{\prime },p) &=&\tilde{\Gamma}_{\mu }^{a}(-p,-p^{\prime
}), \\
\Theta ^{a}(p^{\prime },p) &=&\tilde{\Theta}^{a}(-p,-p^{\prime }), \\
\Omega ^{a}(p^{\prime },p) &=&\tilde{\Omega}^{a}(-p,-p^{\prime }).
\end{eqnarray}
The functions $\Theta ^{a}(p^{\prime },p)$ and $\Omega ^{a}(p^{\prime },p)$ depend on the variables $p^{\prime 2},~p^{2}$ and $q^{2}$. Symmetry (antisymmetry) in the isospin indices implies that vertex function is symmetric (antisymmetric) under the permutation of $p^{\prime 2}$ and $p^{2}$. Functions that are symmetric in isospin describe states with isospin $T=0$ and $2$, whereas those that are antisymmetric describe isospin $T=1$ states.
Similar properties characterize the form factors of $\Gamma _{\mu }^{a}(p^{\prime },p)$.

The vertex functions can be expanded in scalar functions ${\mathcal{F}}_{i \pm}^{a}$ that are symmetric in $p^{\prime 2}$ and $p^{2}$. The lower index $\pm$ indicates the symmetry with respect to permutation of the isospin indices: ${\tilde {\mathcal{F}}}_{i \pm }^{a} = \pm {\mathcal{F}}_{i \pm}^{a} $ ($i=1,2,3$). The expansion takes the form
\begin{eqnarray}
\Gamma _{\mu }^{a}(p^{\prime },p) &=& (p^{\prime }+p)_{\mu }\left( {
\mathcal{F}}_{1-}^{a}+(p^{\prime 2} - p^{2}){\mathcal{F}}_{1+}^{a}\right) + q_{\mu }\left( (p^{\prime 2} - p^{2}){\mathcal{F}}_{2-}^{a}+{\mathcal{F}}_{2+}^{a}\right) ,  \label{29a} \\
\Theta ^{a}(p^{\prime },p)+\Omega ^{a}(p^{\prime },p) &=& {\mathcal{F}}_{3+}^{a}+(p^{\prime 2}-p^{2}){\mathcal{F}}_{3-}^{a}.
\label{29b}
\end{eqnarray}

If there were no isospin symmetry breaking, we could have
${\mathcal{F}}_{i -}^{a} = T^{a}{\mathcal{F}}_{i -}$, as there are no other $SU(2)$ generators, and
${\mathcal{F}}_{i +}^{a} = 0$, implying that %in the mass-shell limit
$\Gamma _{\mu}^{a}(m^{2},m^{2},q^{2})\propto T^{a}$ and $\Theta ^{a}(m^{2},m^{2},q^{2})+\Omega
^{a}(m^{2},m^{2},q^{2})=0$.

The gWTI, Eq.~(\ref{20}), can be split into isospin-symmetric and isospin-antisymmetric parts:
\begin{eqnarray}
-[m^{2}+\frac{1}{2}(\Sigma (p^{\prime })+\Sigma (p)),T^{a}]&=&(p^{\prime}{}^{2}-p^{2})^{2} {\mathcal{F}}_{1+}^{a}
+q^{2} {\mathcal{F}}_{2+}^{a}-{\mathcal{F}}_{3+}^{a},  \label{30a} \\
T^{a}-\frac{1}{2}\{\frac{\Sigma (p^{\prime })-\Sigma (p)}{p^{\prime
}{}^{2}-p^{2}},T^{a}\} &=&{\mathcal{F}}_{1-}^{a}+q^{2}{\mathcal{F}}_{2-}^{a}
- {\mathcal{F}}_{3-}^{a}.  \label{30b}
\end{eqnarray}
where $[,]$ is a commutator and $\{,\}$ is an anticommutator.

Consider some of the consequences of Eq.~(\ref{30a}). For $p^{\prime }=p$%
, $q=0$, we obtain
\begin{equation}
\lbrack m^{2}+\Sigma (p^{2}),T^{a}]={\mathcal{F}}%
_{3+}^{a}(p^{2},p^{2},0).
\label{31}
\end{equation}
In a cyclic basis, the $SU(2)$ generators are equal to $T_{\pm 1}= \mp (T^{1} \pm iT^{2})/\sqrt{2}$ and $T_{0}=T^{3}$. The left-hand side of Eq.~(\ref{31}), when sandwiched between the initial and final states in the pion $\beta$ decay, gives the pion mass difference:
\[
\langle \pi ^{0}|[m^{2}+\Sigma (p^{2}),T_{-1}]|\pi ^{+}\rangle
=m_{f}^{2}-m_{i}^{2} \equiv - \Delta m_{\pi}^{2},
\]
where $m_{\pi ^{0}} = m_{f}$ and $m_{\pi ^{+}} = m_{i}$. In the more general case of $p^{\prime 2} = p^{2}$ and $q\neq 0$, one finds
\begin{equation}
\lbrack m^{2}+\Sigma (p^{2}),T^{a}]=-q^{2}{\mathcal{F}}%
_{2+}^{a}(p^{2},p^{2},q^{2})+{\mathcal{F}}%
_{3+}^{a}(p^{2},p^{2},q^{2}).
\label{32}
\end{equation}
The trace of the commutator of finite-dimensional matrices vanishes, and therefore, the left-hand side of Eqs.~(\ref{31}) and (\ref{32}) describes a state of total isospin 2. %(cf. (\ref{proj})).
The isospin-zero component of Eq.~(\ref{32}) satisfies the equation $q^{2}({\mathcal{F}}_{2+}^{a})^{T=0}-({\mathcal{F}}_{3+}^{a})^{T=0}=0$. At zero momentum transfer,
$({\mathcal{F}}_{3+}^{a})^{T=0}=0$, and the overall normalization of the $T=0$ form factors is not fixed. In the $t$ channel of the pion $\beta $ decay, the state $\pi ^{+}\pi ^{0}$ has the $T^{3}=1$ isospin projection, so the $T=0$ channel is of no interest.

As a consequence of elastic unitarity and analyticity, the $q^{2}$ dependence of the on-mass-shell form factor is determined by the Jost function $D_{J}^{T}(q^{2})$ that can be constructed in terms of the phase shift in the corresponding channel (see, e.g., \cite{Bjor1965}). We consider an $S$-wave because the last three vertices in $\delta \mathcal{L}_{\mathrm{eff}}$ induced by the transformation (\ref{comp}) do not contain derivatives of $\chi^{a}$, and thus, the $S$-wave $T=2$ channel of the pion-pion scattering is relevant. We therefore write
\begin{equation}
({\mathcal{F}}_{3+}^{a}(m_{f}^{2},m_{i}^{2},q^{2}))^{T=2} =
{\mathcal{F}}_{3+}^{a}(m_{f}^{2},m_{i}^{2},0)/D_{J}^{T=2}(q^{2}),
\label{jost}
\end{equation}
where $D_{J}^{T=2}(0)=1$.
The value of ${\mathcal{F}}_{3+}^{a}(m_{f}^{2},m_{i}^{2},0)$ is
proportional to the mass splitting in the pion multiplet. We
restrict ourselves to the first order in $ \Delta m_{\pi}^{2}$. Accordingly,
the physical pion masses in the arguments of the vertex functions can be replaced by a mean value, e.g., $ \mu = (m_f + m_i)/2$,
provided that these functions, such as ${\mathcal{F}}_{2+}^{a}$ and ${\mathcal{F}}_{3+}^{a}$, are already small.
Using Eqs.~(\ref{31}) - (\ref{jost}), we obtain
\begin{eqnarray*}
({\mathcal{F}}_{3+}^{a}(\mu^2,\mu^2,q^{2}))^{T=2}
&=&[m^{2}+\Sigma(\mu^2),T^{a}]/D_{J}^{T=2}(q^{2}),  \label{35} \\
({\mathcal{F}}_{2+}^{a}(\mu^2,\mu^2,q^{2}))^{T=2}
&=&[m^{2}+\Sigma(\mu^2),T^{a}]{\mathcal{F}}_{2+}(\mu^{2},\mu^{2},q^{2}),  \label{36}
\end{eqnarray*}
where
\begin{equation}
{\mathcal{F}}_{2+}(\mu^2,\mu^2,q^{2})=\frac{1/D_{J}^{T=2}(q^{2})-1}{q^{2}}.
\label{37}
\end{equation}

Let us consider the consequences of Eq. (\ref{30b}). Equation (\ref{29a}) shows that ${\mathcal{F}}_{2-}^{a}$ can be evaluated with the required accuracy in the limit of exact isotopic symmetry with
$[m^2 + \Sigma (p),T^{a}] = {\mathcal{F}}_{i +}^{a} = {\mathcal{F}}_{3 \pm}^{a} = 0$ and
${\mathcal{F}}_{1-}^{a}=T^{a}{\mathcal{F}}_{1-}$. On the mass shell,
moreover, $\Sigma (p)=0$, and thus,
\begin{equation}
{\mathcal{F}}_{2-}(\mu^2,\mu^2,q^{2}) = \frac{1-{\mathcal{F}}_{1-}(\mu^2,\mu^2,q^{2})}{q^{2}}.
\label{38}
\end{equation}
To the leading order of the expansion in $\Delta m_{\pi}^{2}$, the form factor ${\mathcal{F}}_{2-}^{a}$ is determined by the isotopic rotation of the electromagnetic form factor (\ref{7}).

The form factors ${\mathcal{F}}_{2\pm}$, as defined by Eqs.~(\ref{37}) and (\ref{38}), are not singular at $q^2 = 0$. For small $q^2$, they are determined by the pion radii.
The on-shell weak vector current is usually parameterized in the form
\begin{equation}
\langle \pi ^{0}(p^{\prime })|\bar{d}\gamma _{\mu }(1-\gamma _{5})u|\pi
^{+}(p)\rangle =\sqrt{2}((p^{\prime }+p)_{\mu }f_{+}+q_{\mu }f_{-}),
\label{1111}
\end{equation}
where $q_{\mu } = (p^{\prime} - p)_{\mu }$.

The exact CVC condition implies
\begin{equation}
f_{-} = 0.
\label{16min}
\end{equation}.

Partial CVC, as follows from the comparison of Eqs.~(\ref{29a}) and (\ref{1111}), yields
\begin{eqnarray}
f_{-} &= &\left( m_{\pi^{0}}^{2}-m_{\pi ^{+}}^{2}\right)
\left( {\mathcal{F}}_{2-}(\mu^{2},\mu^{2},0)+{\mathcal{F}}_{2+}(\mu^{2},\mu^{2},0)\right)  \nonumber \\
&=&\frac{m_{\pi ^{+}}^{2}-m_{\pi ^{0}}^{2}}{6}\left( \left\langle
r^{2}\right\rangle _{v}^{T=1}-\left\langle r^{2}\right\rangle
_{s}^{T=2}\right) .
\label{16}
\end{eqnarray}
The isovector part is recovered from the electromagnetic vertex, whereas the isotensor part is independent of it. Remarkably,
for a dressed vertex, the $W^{\pm }$ boson, being a member of the weak isospin triplet, is coupled to both the strong isospin triplet and the strong isospin quintet.

Equation~(\ref{16}) can also be derived in a simpler and more direct way, analogously to the standard analysis of $K_{\ell 3}$ decays \cite{Trip2012}.

First, we introduce the form factors
\begin{eqnarray}
f_{v}(q^{2}) &=&\mathcal{F}_{1-}(m_{f}^{2},m_{i}^{2},q^{2}),  \label{VV} \\
f_{s}(q^{2}) &=&\mathcal{F}_{1-}(m_{f}^{2},m_{i}^{2},q^{2})
+q^{2}\left(
\mathcal{F}_{2-}(m_{f}^{2},m_{i}^{2},q^{2})+\mathcal{F}%
_{2+}(m_{f}^{2},m_{i}^{2},q^{2})\right) ,  \label{SS}
\end{eqnarray}%
with $f_{+}=$ $f_{v}(0)$ $=$ $f_{s}(0)=1.$ In the $t$ channel, the space-like component of the current is proportional to $f_{v}(q^{2})$, whereas the time-like component is proportional to $f_{s}(q^{2})$.
By applying the $t$-channel unitarity condition for the form factors, we find that $f_{v}$ is determined by the $J=T=1$ scattering amplitude, whereas $f_{s}$ is determined by the $J=0,\;T=2$ scattering amplitude, where $J$ is the angular momentum of the pions. $f_{v}$ and $f_{s}$ can thus be identified as vector and scalar form factors, respectively, which is consistent with the gWTI.

Second, we assume an expansion in $q^{2}$ for both $f_{v}(q^{2})$ and $f_{s}(q^{2}).$ This assumption conveys the analyticity of the form factors. Equation (\ref{SS}) leads to%
\[
1+\frac{1}{6}\langle r^{2}\rangle _{s}^{T=2}q^{2} = 1+\frac{1}{6}%
\langle r^{2}\rangle _{v}^{T=1}q^{2} +\frac{q^{2}}{m_{f}^{2}-m_{i}^{2}%
}f_{-}+O(q^4),
\]%
such that we recover Eq.~(\ref{16}) and, in addition, observe that it is exact for the on-shell case. However, the connection to the CVC condition remains hidden in this simplified approach.

The nature of the form factor $f_{s}$ can also be clarified with the aid of Eq.~(\ref{1111}). Taking the scalar product of both sides of the equation with the vector $q_{\mu}$ yields
\begin{equation}
\langle \pi ^{0}(p^{\prime })|i\partial ^{\mu }\left( \bar{d}\gamma _{\mu
}(1-\gamma _{5})u\right) |\pi ^{+}(p)\rangle =-\Delta m_{\pi }^{2}\sqrt{2}%
f_{s}(q^{2}).
\label{5555}
\end{equation}
On the left-hand side, we obtain the form factor, which is a Lorentz scalar. Equation~(\ref{5555}) is the formal statement of the non-conservation of the lowering component of the weak vector current.

Equation (\ref{16}) can therefore be derived using either the gWTI or the  kinematic decomposition of Eqs.~(4.17) and (4.18).
Because recourse to the gWTI is optional, Eq.~(\ref{16})
is less interesting for testing the gWTI.
We therefore analyze the form factors off the mass shell.

The WTI of exact symmetry implies, to first order in the displacement $p^{\prime 2}-p^{2}$ and for low momentum transfers,
\begin{equation}
f_{-} = - \frac{p^{\prime 2}-p^{2}}{6}\langle r^{2}\rangle _{v}^{T=1}.
\label{on-shell}
\end{equation}

The WTI of broken symmetry implies,
to first order in $\Delta m^2_{\pi}$ and
the displacement $p^{\prime 2}-p^{2}$
and for low momentum transfers,
\begin{equation}
f_{-} =
- \frac{p^{\prime 2}-p^{2}}{6}\langle r^{2}\rangle _{v}^{T=1}
+ \frac{m_{\pi ^{0}}^{2}-m_{\pi ^{+}}^{2}}{6}\langle r^{2}\rangle _{s}^{T=2}.
\label{offshell}
\end{equation}

The $\pi^{+} \to \pi^{0} e^{+} \nu_e \gamma$ decay rate
depends on the near-mass-shell behavior of the form factor $f_{-}$. Any such reaction may serve for testing the gWTI.

For large momentum transfers, the pion radii in Eqs.~(\ref{16}), (\ref{on-shell}) and (\ref{offshell}) should be replaced with the form factors of Eqs.~(\ref{37}) and (\ref{38}).
Expressions~(\ref{16min}), (\ref{16}), (\ref{on-shell}), and (\ref{offshell}) correspond to the various versions of the CVC condition.

The on-shell form factor $\mathcal{F}^{a}_{1 -}$ is independent of both the gauge and the parameterization. By virtue of Eqs.~(\ref{37}) and (\ref{38}), the on-shell form factors $\mathcal{F}^{a}_{2\pm}$ are also uniquely defined. The longitudinal component of the vertex (\ref{29a}) contains the factor $p^{\prime 2} - p^{2}$ in $\mathcal{F}^{a}_{2-}$, whereas $\mathcal{F}^{a}_{2+}$ has smallness of $O(\Delta m^2_{\pi})$. We thus conclude that the longitudinal component of $\Gamma _{\mu }^{a}(p^{\prime },p)$ is uniquely defined in the neighborhood of the mass shell to first order in the pion mass splitting and the displacement $p^{\prime 2}-p^{2}$.
In the neighborhood of the mass shell, the form factor $f_{-}$ thus avoids the general rule \cite{Haag1958,Chis1961,Kame1961,Fear2000,Fear2000a,Bial1970,Hoof1972}
that states that off-shell amplitudes are ambiguous.

The results obtained for the near-mass-shell representation of $f_{-}$ in terms of the physical masses and radii of the pions exhibit explicit independence of the gauge and the parameterization of the pion field.

From a mathematical point of view, one can speak of the uniqueness of the longitudinal component of $\Gamma _{\mu }^{a}(p^{\prime },p)$ in an infinitesimal neighborhood of the mass shell or, equivalently, of the uniqueness of the longitudinal component of $\Gamma _{\mu }^{a}(p^{\prime },p)$ and its first derivatives with respect to $ p^{\prime}_{\nu} $ and $p_{\nu}$ on the mass shell to first order in the pion mass splitting.

The transverse component of the vertex does not exhibit any specific behavior, and thus, we consider $f_{+}$ to be ambiguous off shell.

The generality of the relationships~(\ref{16}) and (\ref{offshell}) is influenced by only the CVC condition at a bare interaction vertex.
This condition is satisfied in the Standard Model,
so any violation of these relationships %(\ref{16}) and (\ref{offshell})
can be interpreted as an indication for new physics at or above the electroweak scale.

%%%%%%%%%%%%%%%%%%%%%%%%%%%%%%%%%%%%%%%%%%%%%%%%%%%%%%%%%%%%%%%%%%%%%%%%%%%%%%%%%%%%%%%%%%%%%%%%%%%%%%%%%%%%%%%%%%%%%%%%%%%%%%%%%%%%%%%%%%%%%%%%%%%%%%
\section{\textbf{Numerical estimates}}
\renewcommand{\theequation}{5.\arabic{equation}}
\setcounter{equation}{0}
%%%%%%%%%%%%%%%%%%%%%%%%%%%%%%%%%%%%%%%%%%%%%%%%%%%%%%%%%%%%%%%%%%%%%%%%%%%%%%%%%%%%%%%%%%%%%%%%%%%%%%%%%%%%%%%%%%%%%%%%%%%%%%%%%%%%%%%%%%%%%%%%%%%%%%

In this section, we employ the dispersion techniques for numerical estimation of the pion radius $\langle r^2 \rangle^{T=2}_s $, entering
the equations~(\ref{16}) and (\ref{offshell}), and of the form factor $f_{-}$.

In the scattering of $T=2$ pions, the inelastic channels do not become relevant until an invariant mass of 1.2 GeV \cite{1111,2222,3333}.
The elastic scattering amplitude is determined by the phase shift through the Jost function:
\begin{equation}
D_{J}^{T}(s)=\exp (-\frac{s}{\pi }\int_{4\mu ^{2}}^{+\infty }\frac{\delta
^{T}(s^{\prime })ds^{\prime }}{s^{\prime }(s^{\prime }-s)}).  \label{B3}
\end{equation}
The unitarity relation for the form factor implies (cf. (\ref{jost})) %(see, e.g., \cite{Bjor1965})
\begin{equation}
({\mathcal{F}}_{3+}(s))^{T}=P(s)/D_{J}^{T}(s),  \label{B4}
\end{equation}%
where $P(s)$ is a finite-degree polynomial. The degree of $P(s)$ can be fixed by the quark-counting rules, which, in our case, take the form ${\mathcal{F}}_{2 \pm}(s)\sim 1/s$ for $s\rightarrow -\infty $. The asymptotes of the Jost function are determined by value of the phase at infinity. The experimental phase is known for energies up to $\sqrt{{s}_{\max }}=2.1$ GeV. At this energy, $\delta (s_{\max})\approx 0$. If one takes $\delta (s)=0$ for $\sqrt{s}>2.1$ GeV, then the quark-counting rules are not satisfied, even for the zero-degree polynomial $P(s)=1$. The minimal required modification is to introduce a hypothetical resonance with mass $m_{X} > \sqrt{{s}_{\max}}$. Such a resonance should have a large width because it is not observed experimentally. Note that in the Veneziano model, there are no resonances in
the $T=2$ channel. The contribution of a large $s$ is suppressed in the
dispersion integral by the inverse of the second power of $s$, and one can
assume that this contribution is not dominant. The contribution of the
large-$s$ region can be evaluated by comparing the contributions of the
intervals $\sqrt{s}=2\mu - 1.5$ GeV and $1.5-2.1$ GeV, for which
experimental data are available. A polynomial $P(s)$ of degree $n$
requires $n+1$ resonances. If there are primitives, then the number of resonances increases. The numbers of resonances and primitives are correlated with the number of Castillejo-Dalitz-Dyson poles \cite{MIK2010}.

Expanding the Jost function in the vicinity of $s=0$, we obtain
\begin{equation}
\langle r^{2}\rangle _{s}^{T}=\frac{6}{\pi }\int_{4\mu ^{2}}^{+\infty }%
\frac{\delta ^{T}(s^{\prime })}{s^{\prime 2}}ds^{\prime }.
\label{B5}
\end{equation}%
For numerical estimates, the upper limit of the integral is replaced by
$s_{\max}$; the scattering phase shift between the experimental points is interpolated linearly. The results are shown in Table 1 for the data sets from Refs.~\cite{1111,2222,3333}
A further refinement of the estimate %of $\langle r^{2}\rangle _{s}^{T=2}$
can be done using more theoretical inputs from
semi-phenomenological models of pion-pion scattering \cite{Xiem2014}
and lattice QCD. \cite{Bean2012}

The negative ``mean square radius'' occurs because of the negative $t$-channel scattering phase, which is a signature of repulsion.
In the general case,
the mean square radius is defined as
the first expansion coefficient
of the corresponding form factor
in powers of $q^2 $, and
the sign of this expansion coefficient is not fixed a priori.
Vector boson exchange produces repulsive forces between identical particles.
The $T = 2$ channel with $T_{3}=2$ contains a pair of positively charged pions,
and thus, the repulsive nature of the phase can be attributed to the dominance of the $t$-channel $\rho$-meson exchange between the pions.

Equation~(\ref{B5}) has a variety of applications. In the zero-width approximation with $\delta^{T=1} (s)=\pi \theta (s-m_{\rho }^{2})$, where $m_{\rho }$ is the mass of the $\rho$ meson, one arrives at the well-known expression for the pion charge radius, $\langle r^{2}\rangle _{v}^{T=1}=6/m_{\rho }^{2}$. In the $T=1$ channel with an attraction, $\langle r^{2}\rangle _{v}^{T=1}$ becomes positive and approaches the experimental value, thereby supporting the vector meson dominance model.
The pion scalar radius can be determined using the data for pion-pion scattering in the $T=0$ channel, yielding $\left\langle r^{2}\right\rangle_{s}^{T=0}=0.61 \pm 0.04\,\,\mathrm{fm}^{\mathrm{2}}$ \cite{Cola2001}. The positive value of
$\left\langle r^{2}\right\rangle _{s}^{T=0}$ reflects the dominance of the attractive interaction, which is consistent with the existence of the $\sigma$ meson.

In the dispersion theory, the contribution from the high energy part of the dispersion integrals is considered to be the important source of theoretical uncertainties. The representation~(\ref{B3}) is justified in an approximation of neglecting all the inelastic channels.
In our case, the elasticity holds for $\sqrt{s}=2\mu - 1.2$ GeV,\cite{1111,2222,3333}
so the dispersion integral from this region accurately assesses the contribution to the pion radius.
With increasing the energy above 1.2 GeV, the parameter of inelasticity increases smoothly up to 0.5 for $\sqrt{s}=2.1$ GeV.
The part of the integral (\ref{B5}) corresponding to this region gives therefore an approximate estimate of the contribution.
Using results of Ref.~\cite{2222}, we find that this contribution is numerically small, and equals only $- 0.007$ fm$^2$.
The uncertainty associated with higher energies $\sqrt{s} > 2.1$ GeV
can be estimated by assuming the existence of a hypothetical resonance $X$ with mass $m_ {X} > 2.1$ GeV.
In the zero-width approximation, the error equals $ 6/m_{X}^{2} \sim 0.03$ fm$^{2}$ for $ m_ {X} = 3 $ GeV, which is comparable to the experimental uncertainty of the contribution from the interval $\sqrt{s}=2\mu - 1.2$ GeV.

%%%%%%%%%%%%%%%%%%%%%%%%%%%%%%%%%%%%%%%%%%%%%%%%%%%%%\simeq
%%%%%%%%%%%%%%%%%%%%%%%%%%%%%%%%%%%%%%%%%%%%%%%%%%%%%5
%%%%%%%%%%%%%%%%%%%%%%%%%%%%%%%%%%%%%%%%%%%%%%%%%%%%%55

\begin{table}[t]
\caption{Lorentz scalar isospin $T=2$ mean square radius of the pion, as determined using Eq.~(\ref{B5}). The phase-shift analyses of Refs.~\cite{1111,2222,3333} are used; (a) and (b) denote non-equivalent datasets of scattering phases. $\sqrt{{s}_{\max}}$ is the maximum energy up to which the phase analysis is performed.}
\label{tab:1}
\centering
\vspace{2pt}
%\par
\begin{tabular}{|c|c|c|}
\hline\hline
$\langle r^2 \rangle^{T=2}_s $ [fm$^2$] & $\sqrt{{s}_{\max}}$ [GeV] & Ref. \\
\hline\hline
$-0.09 \pm 0.05$ & $1.4$ & \cite{1111} \\
$-0.10 \pm 0.01$ & $2.1$ & \cite{2222}~(a) \\
$-0.10 \pm 0.03$ & $2.1$ & \cite{2222}~(b) \\
$-0.11 \pm 0.02$ & $1.5$ & \cite{3333}~(a) \\
$-0.13 \pm 0.01$ & $1.5$ & \cite{3333}~(b) \\ \hline\hline
\end{tabular}%
\end{table}

%%%%%%%%%%%%%%%%%%%%%%%%%%%%%%%%%%%%%%%%%%%%%%%%%%%%%555
%%%%%%%%%%%%%%%%%%%%%%%%%%%%%%%%%%%%%%%%%%%%%%%%%%%%%555
%%%%%%%%%%%%%%%%%%%%%%%%%%%%%%%%%%%%%%%%%%%%%%%%%%%%%555

Experimental value of the $\pi ^{+}$ charge radius equals
$\left\langle r^{2}\right\rangle _{v}^{T=1}=\left( 0.672\pm 0.008\,\,\mathrm{%
fm}\right) ^{\mathrm{2}}$ \cite{PDG}. Using the value $\left\langle
r^{2}\right\rangle _{s}^{T=2}=-0.10\pm 0.03\pm 0.03\,\,\mathrm{fm}^{\mathrm{2}}$, we
obtain with the help of Eq. (\ref{16})
\begin{equation}
f_{-}=(2.97\pm 0.25)\times 10^{-3},
\end{equation}
which is a factor of two greater than the light-front quark model
prediction \cite{Jaus1999}.
The estimated error corresponds to
the experimental errors in the $\pi\pi$-scattering phases %($\pm 0.03$ fm$^2$)
and the uncertainty of the high energy part of the integral representation (\ref{B5}). %($\pm 0.03$ fm$^2$). %($\sim  6/m_{X}^2$).
%The contribution to $f_{-}$ of the scalar charge radius is five times smaller than the vector charge radius.

Contribution of the longitudinal form factor
to the $\pi ^{+}\rightarrow \pi ^{0}e^{+}\nu _{e}$ decay rate can be estimated to give
$\Delta B/B = -0.94 \times 10^{-3}f_{-}$;
the additional small factor in $f_{-}$ arises for
kinematic reasons. The experimental error in $B$ is 0.6\% \cite{PDG}. We
thus reaffirm earlier conclusions that $f_{-}$ is currently beyond the capabilities of the experimental study.
The possibility of measuring the longitudinal weak vector current in the neutron $\beta$-decay \cite{Abel2008}, muon capture \cite{Gazi2008} and in the $\tau ^{-}\rightarrow \pi ^{-}\pi ^{0}\nu _{\tau }$ and $\tau ^{-}\rightarrow K^{-}K^{0}\nu _{\tau}$ decays is, perhaps, more promising.

A favorable experimental situation exists in the $K_{\ell 3}$ decays.
The large difference in the masses of $\pi$- and $K$-mesons hampers the near-mass-shell expansion, so the $K_{\ell 3}$ decays seem to be not suitable for the measuring $f_{-}$ off-shell and testing the gWTI. At the same time the dispersion techniques allowing to determine $f_{-}$ on-shell have found in these decays a successful application \cite{Bern2006,Bern2009}.

The dominant part of $\Delta m_{\pi}^2$ is electromagnetic in origin \cite{DasT1967,Gass1984,Urec1995}.
Small mass difference between up and down quarks leads to the $\pi^{0}-\eta$ mixing, which also contributes to $\Delta m^2_{\pi}$.
Equation~(\ref{16}) describes therefore the electromagnetic contribution of order $O(\alpha )$ to the longitudinal weak vector current and the small $\pi^{0}-\eta$ mixing effect.

$\chi$PT has been very successful in describing properties
of light mesons in the non-perturbative regime \cite{Gass1984,Urec1995,Gass1985,SCHR2002}.
Predictions for the $\pi \pi$-scattering lengths, first obtained by Weinberg \cite{Wein1966} and improved by the development of $\chi$PT,
agree with experiment to a high degree of accuracy. \cite{Gass2009,Cola2009}
The pion radii $\langle r^2 \rangle^{T=0}_s $ and $\langle r^2 \rangle^{T=1}_v $ are also well understood \cite{Cola2001,Gass1984,Gass1991,Bijn1998},
while the pion radius $\langle r^2 \rangle^{T=2}_s $ %, entering Eqs.~(\ref{16}) and (\ref{offshell}),
has not been discussed so far.
The embedding of isotopic symmetry breaking into the $\chi$PT framework on line with Refs.~\cite{Pich2003,Pich2003a,Urec1995,Ciri2007}
could allow to also describe the near-mass-shell form factor $f_{-}$, starting from the first principles of QCD.

%%%%%%%%%%%%%%%%%%%%%%%%%%%%%%%%%%%%%%%%%%%%%%%%%%%%%%%%%%%%%%%%%%%%%%%%%%%%%%%%%%%%%%%%%%%%%%%%%%%%%%%%%%%%%%%%%%%%%%%%%%%%%%%%%%%%%%%%%%%%%%%%%%%%%%
\section{Conclusion}
\renewcommand{\theequation}{6.\arabic{equation}}
\setcounter{equation}{0}
%%%%%%%%%%%%%%%%%%%%%%%%%%%%%%%%%%%%%%%%%%%%%%%%%%%%%%%%%%%%%%%%%%%%%%%%%%%%%%%%%%%%%%%%%%%%%%%%%%%%%%%%%%%%%%%%%%%%%%%%%%%%%%%%%%%%%%%%%%%%%%%%%%%%%%

In this paper, the longitudinal component of the weak vector current in the $\pi_{e3}$ decay was calculated. This component arises due to the violation of isotopic symmetry by the pion masses and the electromagnetic and weak interactions. A generalization of the WTI was derived in the pion sector to account for the isotopic symmetry breaking.
It was shown that the isovector $ T = 1 $ part of the current can be reconstructed through isotopic rotation of the off-shell pion electromagnetic form factors, whereas the isotensor $ T = 2 $ part has no analogs but rather is uniquely determined by the gWTI combined with
the elastic unitarity, analyticity, and the pion-pion scattering data.
Because of the low momentum transfer, the on-shell form factor
is proportional to the pion mass difference and the difference of the vector-isovector and scalar-isotensor pion radii. The relationship between the near-mass-shell form factor and the pion radii, Eq.~(\ref{offshell}), is a consequence of the CVC condition
extended beyond the approximation of exact isotopic symmetry.

The various versions of the CVC condition are distinguished depending on whether the isotopic symmetry is exact or broken and whether the outer legs in the vertex are on or off shell.
The corresponding predictions for the longitudinal form factor $f_{-}$
are given in Eqs.~(\ref{16min}), (\ref{16}), (\ref{on-shell}), and (\ref{offshell}).
In the case of the exact isotopic symmetry, the bare and dressed weak vertices are pure isospin triplets; the CVC and the WTI hold on and off shell, respectively. In the case of the broken isotopic symmetry, the dressed weak vertex is no longer a pure isospin triplet; partial CVC and the gWTI hold on and off shell.

Favorable conditions for testing the gWTI by measuring the longitudinal weak vector current appear to exist
in the $\tau ^{-}\rightarrow \pi ^{-}\pi ^{0}\nu _{\tau } \gamma$ and $\tau ^{-}\rightarrow K^{-}K^{0}\nu _{\tau} \gamma$ decays
with soft photon emission
because of
the involvement of the near-mass-shell form factors,
the relatively large momentum transfer in the weak vertex, and
the small mass differences within the pion and kaon multiplets.

A non-trivial consequence of partial CVC occurs in the off-shell case,
in which the vertex in general is not gauge invariant and depends on the parameterization of the pion field.
The only exception is the longitudinal component of the vertex in the neighborhood of the mass shell.
To first order in the pion mass splitting and the displacement from the mass shell, the longitudinal component is independent of both the gauge
and the parameterization, and therefore, the near-mass-shell form factor $f_{-}$ appears to be a unique object whose properties
are unambiguously determined by the partial CVC condition (the gWTI) and also fundamentally allow for experimental verification.

%\vspace{-1pt}

\section*{Acknowledgments}
This work was supported by RFBR Grant No.~13-02-01442, Grant~No.~HLP-2015-18~of Heisenberg-Landau~Program,
and Grant No.~3172.2012.2 for Leading Scientific Schools of Russian Federation.

\newpage
%\vskip 20pt

\renewcommand{\thesection}{}

\end{document}